\documentclass{article}
\usepackage{arxiv}

\usepackage[utf8]{inputenc} 
\usepackage[T1]{fontenc}    
\usepackage{hyperref}       
\usepackage{url}            
\usepackage{booktabs}       
\usepackage{amsfonts}       
\usepackage{nicefrac}       
\usepackage{microtype}      
\usepackage{lipsum}
\usepackage{graphicx}
\usepackage{subcaption} 
\usepackage{amsmath}
\usepackage{natbib}
\usepackage{braket}
\usepackage{url}
\usepackage{tabularx}
\usepackage{booktabs}
\usepackage{caption}
\usepackage[linesnumbered,ruled,vlined]{algorithm2e}
\usepackage{bm}
\usepackage{multicol}

\title{Large-scale portfolio optimization using Pauli Correlation Encoding}

\author{
Vicente P. Soloviev \\
Fujitsu Research of Europe Ltd. \\
Madrid, Spain\\
\texttt{vicente.perezsoloviev@fujitsu.com} \\
   \And
Michal Krompiec \\
Fujitsu Research of Europe Ltd. \\
Slough, UK \\
\texttt{michal.krompiec@fujitsu.com} \\
}

\begin{document}
\twocolumn[
\maketitle
\begin{abstract}
Portfolio optimization is a cornerstone of financial decision-making, traditionally relying on classical algorithms to balance risk and return. Recent advances in quantum computing offer a promising alternative, leveraging quantum algorithms to efficiently explore complex solution spaces and potentially outperform classical methods in high-dimensional settings. However, conventional quantum approaches typically assume a one-to-one correspondence between qubits and variables (e.g. financial assets), which severely limits the applicability of gate-based quantum systems due to current hardware constraints. As a result, only quantum annealing-like methods have been used in realistic scenarios. In this work, we show how a gate-based variational quantum algorithm can be applied to a real-world portfolio optimization problem by assigning multiple variables per qubit, using the Pauli Correlation Encoding algorithm. Specifically, we address a problem involving over 250 variables, where the market graph representing a real stock market is iteratively partitioned into sub-portfolios of highly correlated assets. This approach enables improved scalability compared to traditional variational methods and opens new possibilities for quantum-enhanced financial applications.
\end{abstract}

\vspace{1em}
\vspace{2em}
]

\section{Introduction}
Portfolio optimization has become a trending problem in financial decision-making, which involves optimal allocation of capital among a set of financial assets to achieve specific investment objectives. It usually consists of maximizing expected return while minimizing the risk \citep{markowitz1952portfolio}. However, as real-world constraints and the dimensionality of financial markets increase, classical optimization approaches often encounter computational bottlenecks due to the complexity of the problem. Recent advances in quantum computing have opened new avenues for addressing such combinatorial optimization challenges \citep{abbas2024challenges}. 

Recent gate-based approaches for portfolio optimization include Quantum Approximate Optimization Algorithm (QAOA) \citep{farhi2014quantum, kerenidis2019quantum, uotila2025higher, huot2024enhancing, stopfer2025quantum} and Variational Quantum Eigensolver \citep{peruzzo2014variational, wang2025variational, wang2025achieving}. However, the main limitation of these approaches is the scalability to real use cases. Specifically, the number of qubits required ($n$) increases linearly with the number of assets ($m$), i.e., $\mathcal{O}(n)$, which quickly exceeds the capacity of current quantum devices as $m$ grows. To mitigate this, recent research has explored circuit cutting, which partition large quantum circuits into smaller, more manageable subcircuits that can be executed independently and recombined classically \citep{soloviev2025scaling}. Nevertheless, even with these advancements, the largest portfolio optimization problems addressed to date remain limited to fewer than $m=100$ assets, underscoring the ongoing challenge of scaling gate-based quantum algorithms for practical financial optimization tasks. 

In addition to gate-based quantum algorithms, alternative approaches such as quantum annealing and digital annealer \citep{sakuler2025real, phillipson2021portfolio} have been explored for portfolio optimization, particularly for large-scale and combinatorial instances. They both utilize specialized hardware or algorithms to mimic the annealing process and solve similar Quadratic Unconstrained Binary Optimization (QUBO) formulations at scale. These methods have demonstrated the ability to handle portfolio problems involving hundreds or even thousands of assets, offering a level of scalability that currently surpasses gate-based quantum devices. However, despite their practical utility, neither quantum annealing nor digital annealing has yet exhibited a clear quantum advantage over state-of-the-art classical optimization algorithms\cite{stopfer2025quantum}.

In this paper we use a recently developed qubit encoding\cite{sciorilli2025towards} in which each of the qubits in the quantum system encodes more than one variable of the optimization problem. By using this encoding in combination with a variational approach we drastically reduce the required number of qubits, and solve a portfolio optimization problem over a market graph with more than $m=250$ variables (number of stock assets). We show an extensive benchmarking against QAOA (for small instances) and a classical optimizer (for large instances), following an iterative bipartition strategy.

The article is structured as follows: Section~\ref{sec_background} describes the main concepts regarding portfolio optimization and the qubit encoding used in the approach; Section~\ref{sec_method_results} describes the followed methodology and results found; Section~\ref{sec_conclusions} rounds the paper off with some further conclusions and future research lines.

\section{Background}
\label{sec_background}

\subsection{Portfolio optimization}
\label{sec_portfolio_optimization}

Portfolio optimization is a key component of contemporary quantitative finance, which focuses on allocating capital across a group of assets in order to strike the best possible balance between risk and projected return. Markowitz \cite{markowitz1952portfolio} developed the traditional mean-variance paradigm, which represents a portfolio's expected return as the weighted sum of the returns of its individual assets:

\begin{equation}
\label{eq_return}
\mu_p = \sum_{i=1}^N w_i \mu_i
\end{equation}

where $\mu_p$ is the expected portfolio return, $w_i$ is the weight of asset $i$ in the portfolio, $\mu_i$ is the expected return of asset $i$, and $N$ is the total number of assets.

Risk is typically quantified by the portfolio variance, which incorporates both the variances of individual assets and their pairwise correlations:

\begin{equation}
\sigma_p^2 = \sum_{i=1}^N \sum_{j=1}^N w_i w_j \sigma_{ij}
\end{equation}

where $\sigma_{ij}$ is the covariance between assets $i$ and $j$. The covariance can be further expressed in terms of the correlation coefficient $\rho_{ij}$ and standard deviations $\sigma_i, \sigma_j$:

\begin{equation}
\sigma_{ij} = \rho_{ij} \sigma_i \sigma_j
\end{equation}

The correlation coefficient $\rho_{ij}$ measures the linear relationship between the returns of assets $i$ and $j$:

\begin{equation}
\label{eq_pearson_pair}
\rho_{ij} = \frac{\text{Cov}(r_i, r_j)}{\sigma_i \sigma_j}
\end{equation}

where $r_i$ and $r_j$ are the returns of assets $i$ and $j$, respectively.

In addition to the mean-variance framework, a widely used metric for evaluating portfolio performance is the Sharpe ratio, introduced by Sharpe (1966). The Sharpe ratio measures the risk-adjusted return of a portfolio by comparing its excess return over a risk-free rate to its standard deviation:

\begin{equation} \label{eq_sharpe} \text{Sharpe Ratio} = \frac{\mu_p - r_f}{\sigma_p} \end{equation}
where $\mu_p$ is the expected portfolio return, $r_f$ is the risk-free rate, and $\sigma_p$ is the standard deviation of portfolio returns. A higher Sharpe ratio indicates a more favorable risk-return trade-off, making it a key criterion in portfolio selection and optimization. Unlike raw return metrics, the Sharpe ratio accounts for volatility, thus enabling more robust comparisons across strategies with differing risk profiles.

This metric is particularly relevant in the context of this study, where multiple optimization strategies are benchmarked not only by their absolute returns but also by their ability to manage risk effectively. In subsequent sections, we use the Sharpe ratio to evaluate the performance of portfolios.

Several real-world limitations, including minimum/maximum position sizes, cardinality constraints, and transaction costs, affect portfolio optimization in practice. As a result, the problem is frequently non-convex and computationally difficult. In the last decades, several approaches directly address the graph representation of the stock market to design efficient portfolios. Thus, a market graph is constructed by representing the financial assets as nodes in the graph and the conditional relationships, such as $\rho_{ij}$ or $\sigma_{ij}$, as edges. 

\begin{figure}[h]
    \centering
    \includegraphics[width=\linewidth]{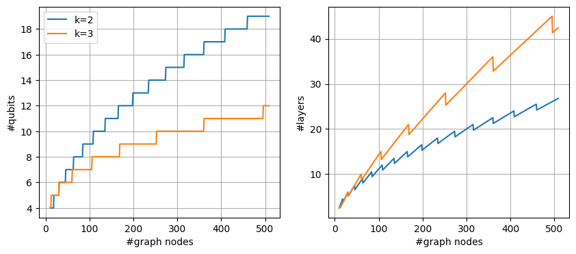}
    \caption{Number of qubits, and layers as a function of the number of variables in the problem to be solved}
    \label{fig_scalability_pce}
\end{figure}

\subsection{Pauli Correlation Encoding}
\label{sec_pce}

\begin{figure*}[t]
\centering
\begin{subfigure}[b]{0.25\textwidth}
\centering
\includegraphics[width=\textwidth]{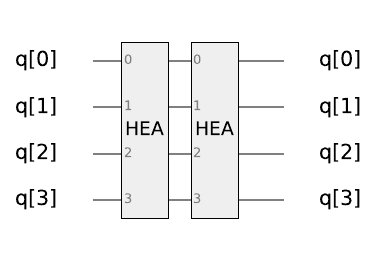}
\caption{HEA}
\label{fig_ansatz_general}
\end{subfigure}
\begin{subfigure}[b]{0.73\textwidth}
\centering
\includegraphics[width=\textwidth]{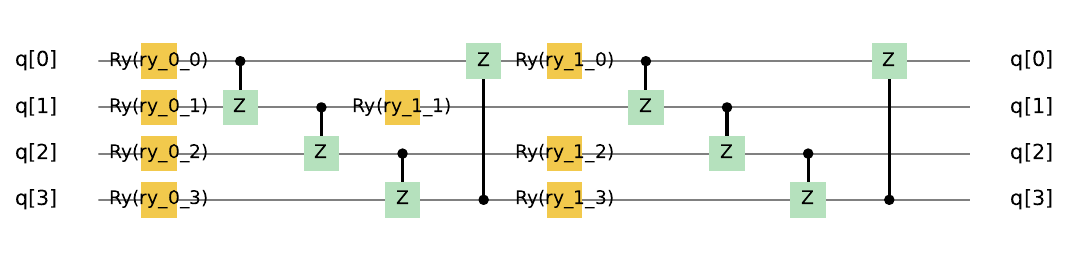}
\caption{Decomposed visualization}
\label{fig_ansatz_decomposed}
\end{subfigure}
\caption{HEA built with a linear entangling structure and CZ gates, sued in PCE strategy approach where $m = 10$, $n=4$, $k=2$ and $p=2$.}
\label{fig_ansatz}
\end{figure*}

Pauli Correlation Encoding (PCE) \citep{sciorilli2025towards} introduces a novel approach for embedding more than one optimization variable per qubit in the system. Specifically, the number of binary variables ($m$) using $n$ qubits are encoded as $\mathcal{O}(n^k)$, where $k$ is the order chosen by the user. This encoding is achieved by using a sign function over the Pauli-matrix correlations across multiple qubits ($k$). Thus, each binary variable is defined as:
\begin{equation}
\label{eq_sgn_func}
    x_i = \text{sgn}(\braket{\Pi_{i}}) \hspace{2em} \forall i \in m,
\end{equation}
where sgn is a sign function, and $\braket{\Pi_{i}} = \bra{\Psi} \Pi_{i} \ket{\Psi}$ is the expectation value of $\Pi_{i}$, over a quantum system $\Psi$. The definition of the Pauli strings ($\Pi_{i}$) depends on the order ($k$) chosen by the user. The seminal paper suggests that a quadratic order will return better results when $m$ is small, and cubic order will outperform the former for an increasing number of variables. Then, we define $\Pi^k = \{\Pi_1^k, \dots, \Pi_m^k\}$ where each $\Pi_i^k$ is a permutation of either $X^k \otimes I^{n-k}$, $Y^k \otimes I^{n-k}$ or $Z^k \otimes I^{n-k}$, which are mutually commuting.

By using this encoding, the number of optimization variables improves upon the one-hot-encoding used in traditional approaches such as QAOA (equivalent to $k=1$). Figure~\ref{fig_scalability_pce}(left) shows the scaling of $n$ as a function of $m$ and $k$ where $k=3$ shows an improvement for larger $m$. 

Another differentiating feature of this approach against other quantum variational approaches is that the quantum circuit used during the runtime is not dependent on the target problem. Then, the complexity of the original problem will not increase the depth of the circuit, as in QAOA, whose depth depends on the number of edges or quadratic terms in the problem. In this paper, we will use the Hardware Efficient Ansatz \citep{kandala2017hardware} circuit. 
The required number of layers ($p$) is defined as $p=\lfloor m/n \rfloor$ (Figure~\ref{fig_scalability_pce}(right)).
Figure~\ref{fig_ansatz} shows a visual representation of the HEA where $m = 10$, $n=4$, $k=2$ and $p=2$.

The variational approach consists in optimizing the parameters according to a cost function that reflect the optimization task to be solved. The seminal PCE paper\cite{sciorilli2025towards} proposes the following function:
\begin{equation}
\label{eq_loss}
    \mathcal{L} = \sum_{(i, j) \in E} w_{ij} \; \tanh{(\alpha \braket{\Pi_{i}})} \; \tanh{(\alpha \braket{\Pi_{j}})} + \mathcal{L}^{reg},
\end{equation}
where $E$ is the number of edges in the original graph, $\alpha$ usually tuned to $\alpha = n^{\lfloor k/2 \rfloor}$, and $\mathcal{L}^{reg}$ is a regularization term.

\section{Method and results}
\label{sec_method_results}

\subsection{Market graph representation}
\label{sec_market_graph_rep}
As explained in Section~\ref{sec_portfolio_optimization}, a market graph is a representation of a real stock market. In this paper, we work with real data from the S\&P 500 stock market\footnote{\url{https://www.kaggle.com/datasets/camnugent/sandp500}} containing 5 years of data. Since we perform the benchmark for various sizes of the portfolio, the experiment with $m$ assets will use the first $m$ columns of the dataset. Thus, the results for bigger $m$ will partially overlap with the results for smaller subsets of assets.

\begin{figure}[h]
\centering
\begin{subfigure}[b]{0.15\textwidth}
\centering
\includegraphics[width=\textwidth]{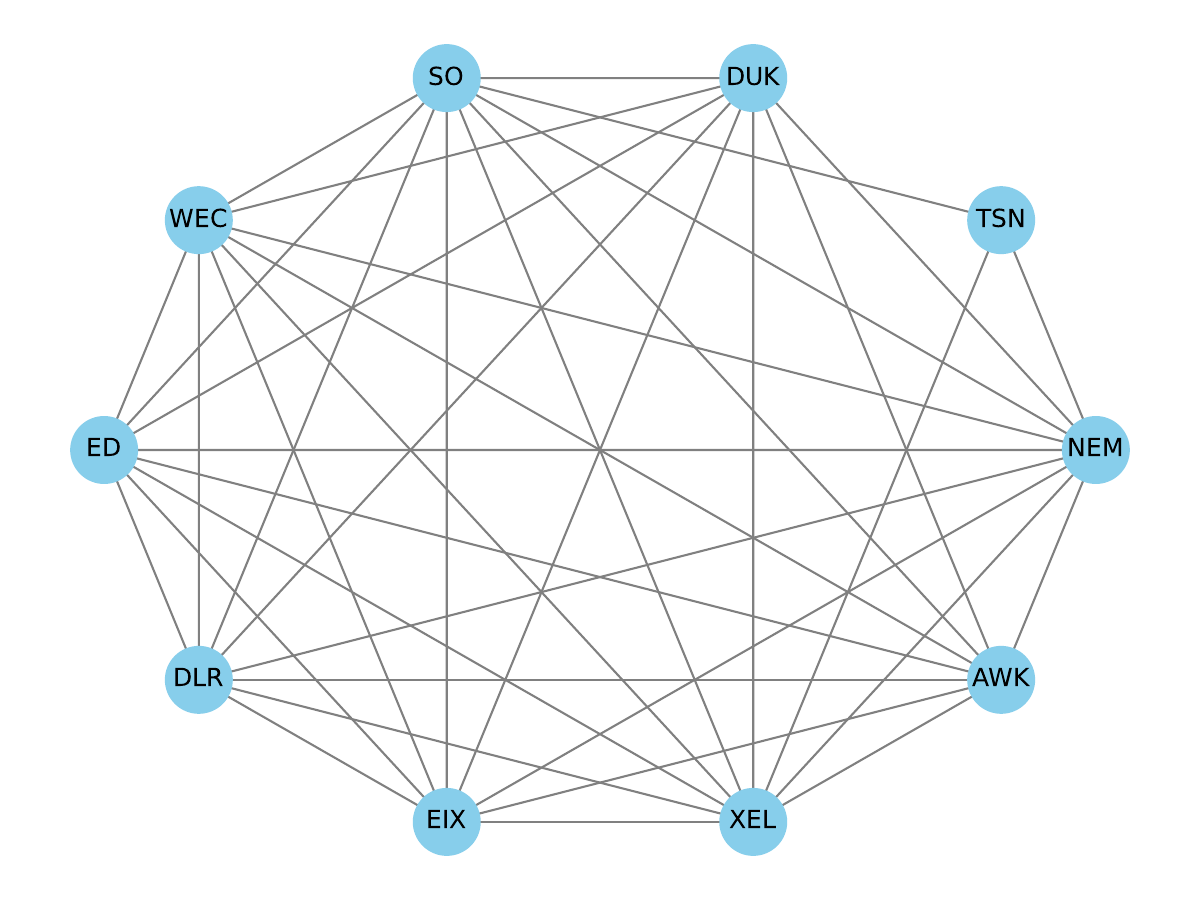}
\caption{$m=10$}
\label{fig_graph_10_plain}
\end{subfigure}
\begin{subfigure}[b]{0.15\textwidth}
\centering
\includegraphics[width=\textwidth]{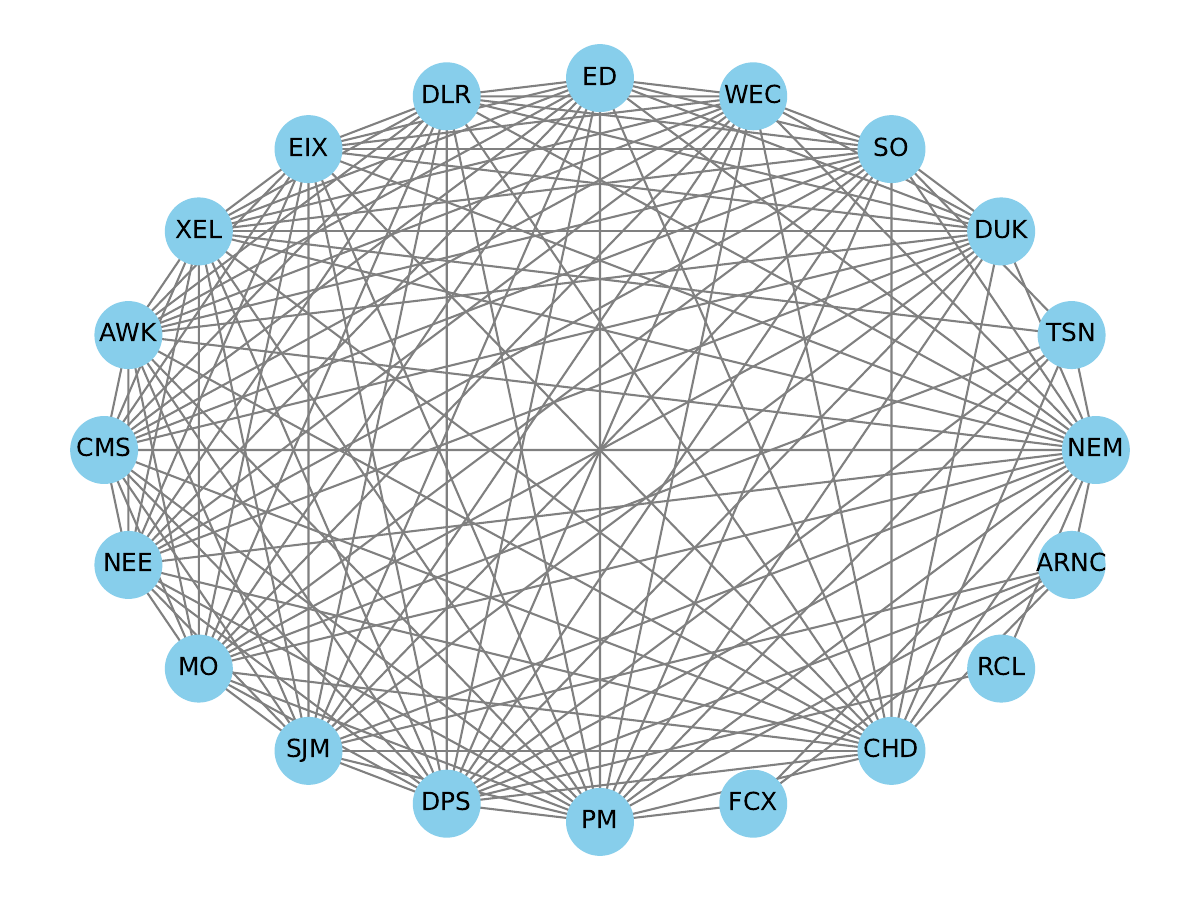}
\caption{$m=20$}
\label{fig_graph_20_plain}
\end{subfigure}
\begin{subfigure}[b]{0.15\textwidth}
\centering
\includegraphics[width=\textwidth]{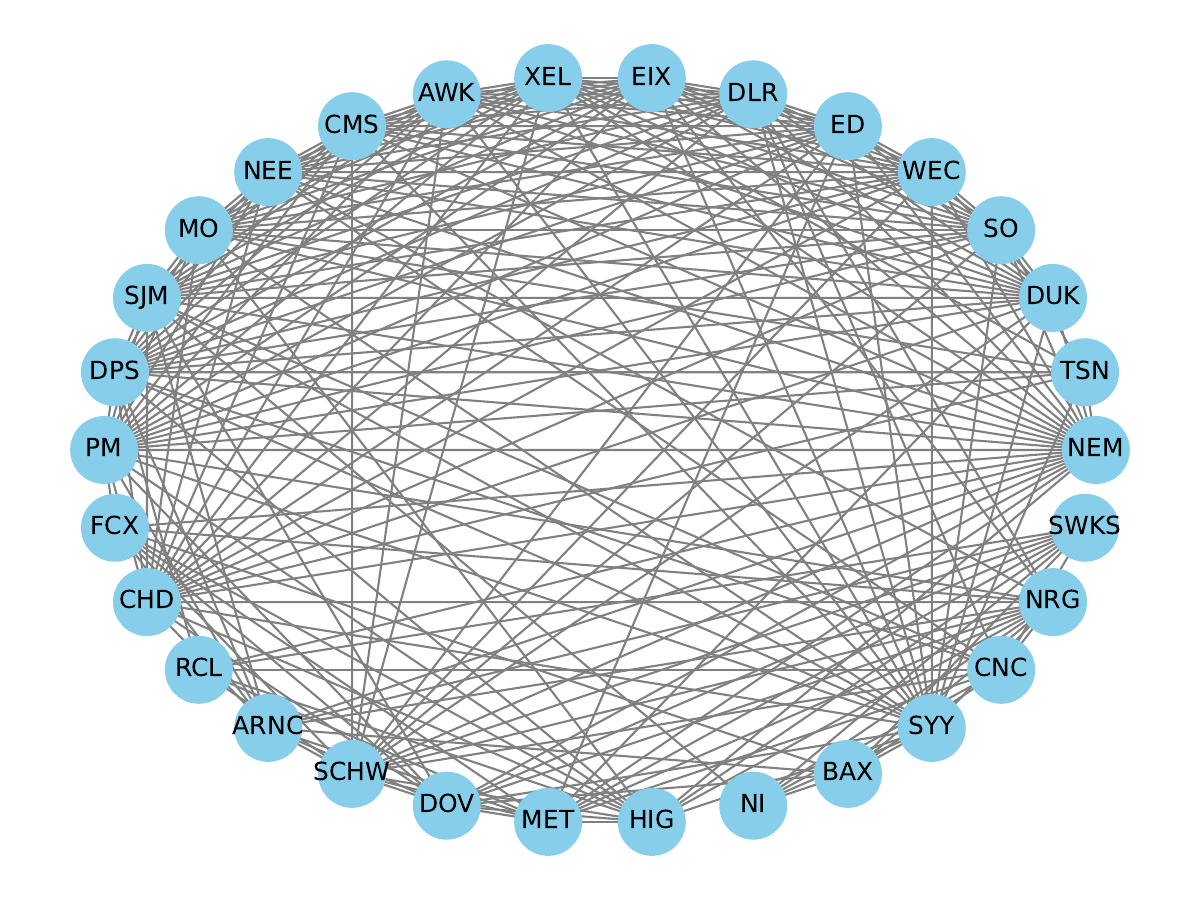}
\caption{$m=30$}
\label{fig_graph_30_plain}
\end{subfigure}
\caption{Market graph with $m$ nodes where nodes represent stock assets.}
\label{fig_graph_10_20_30_plain}
\end{figure}

The market graph $\mathcal{G}$ assigns each asset to a node and contains an edge between nodes $i$ and $j$ only if their pairwise Pearson correlation ($\rho_{ij}$) is above a certain threshold $\lambda$, and no edge otherwise. 
It has been widely studied that an all-to-all graph connectivity in $\mathcal{G}$ does not represent the real behavior of the stock market \citep{chmielewski2020network, namaki2011network}. Thus,
\begin{equation}
\label{eq_weights}
    w_{ij} = 
    \begin{cases}
        1 - |\rho_{ij}| & \text{if} |\rho_{ij}| > \lambda  \\
        \emptyset       & \text{otherwise} $ $
     \end{cases},
\end{equation}
where $|\cdot|$ is the absolute value and $\emptyset$ means that no edge is drawn.

\begin{table*}[ht]
    \centering
    \begin{tabularx}{\textwidth}{@{}X|X|X|X|X@{}}
        \toprule
        \textbf{\# Nodes} & \textbf{\# Edges} & \textbf{Density} & \textbf{Avg. Degree} & \textbf{Clustering Coeff.} \\
        \midrule
        10      & 39      & 0.867   & 7.80        & 0.95              \\
        20      & 137     & 0.721   & 13.7        & 0.92              \\
        30      & 230     & 0.528   & 15.3        & 0.70              \\
        50      & 714     & 0.560   & 28.0        & 0.77              \\
        100     & 3154    & 0.630   & 63.08       & 0.80              \\
        150     & 7606    & 0.680   & 101.4       & 0.82              \\
        200     & 14163   & 0.710   & 141.63      & 0.85              \\
        250     & 20948   & 0.660   & 166.91      & 0.83              \\
        \bottomrule
    \end{tabularx}
    \vspace{1em}
    \caption{Market graph characteristics across different graph sizes}
    \label{tab_graph_info}
\end{table*}

\begin{figure*}[h]
\centering
\begin{subfigure}[b]{0.3\textwidth}
\centering
\includegraphics[width=\textwidth]{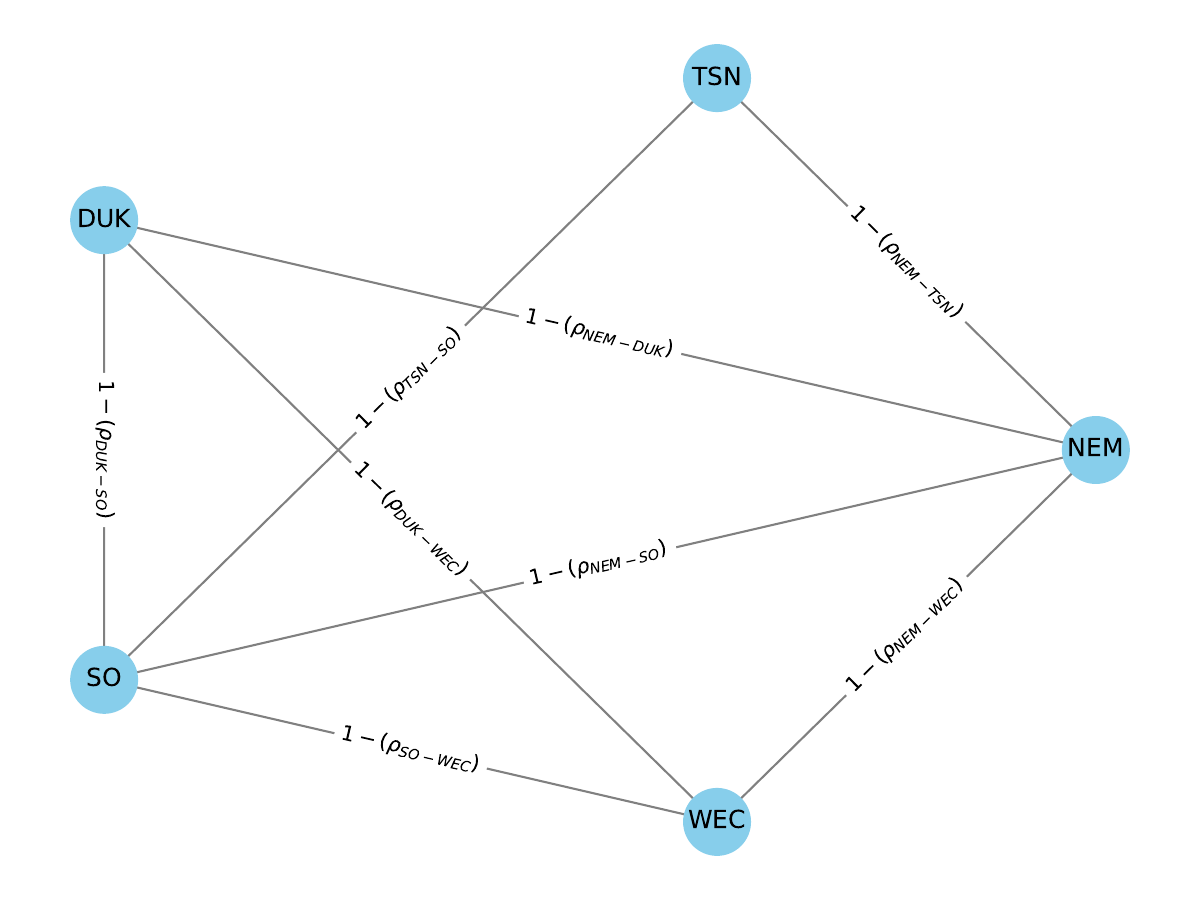}
\caption{Market graph ($m=5$)}
\label{fig_graph_5_plain}
\end{subfigure}
\begin{subfigure}[b]{0.33\textwidth}
\centering
\includegraphics[width=\textwidth]{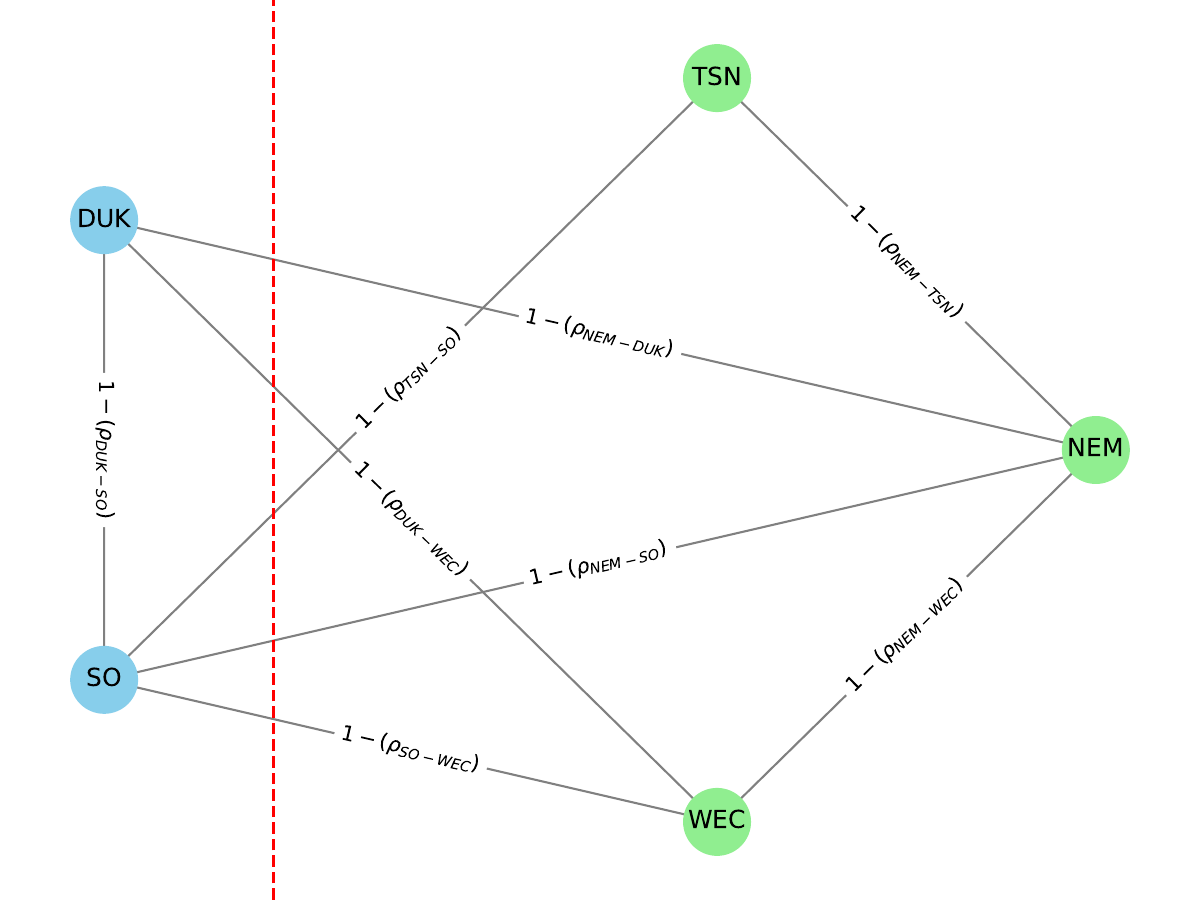}
\caption{Bipartition ($n_{splits} = 1$)}
\label{fig_graph_5_bipart}
\end{subfigure}
\begin{subfigure}[b]{0.33\textwidth}
\centering
\includegraphics[width=\textwidth]{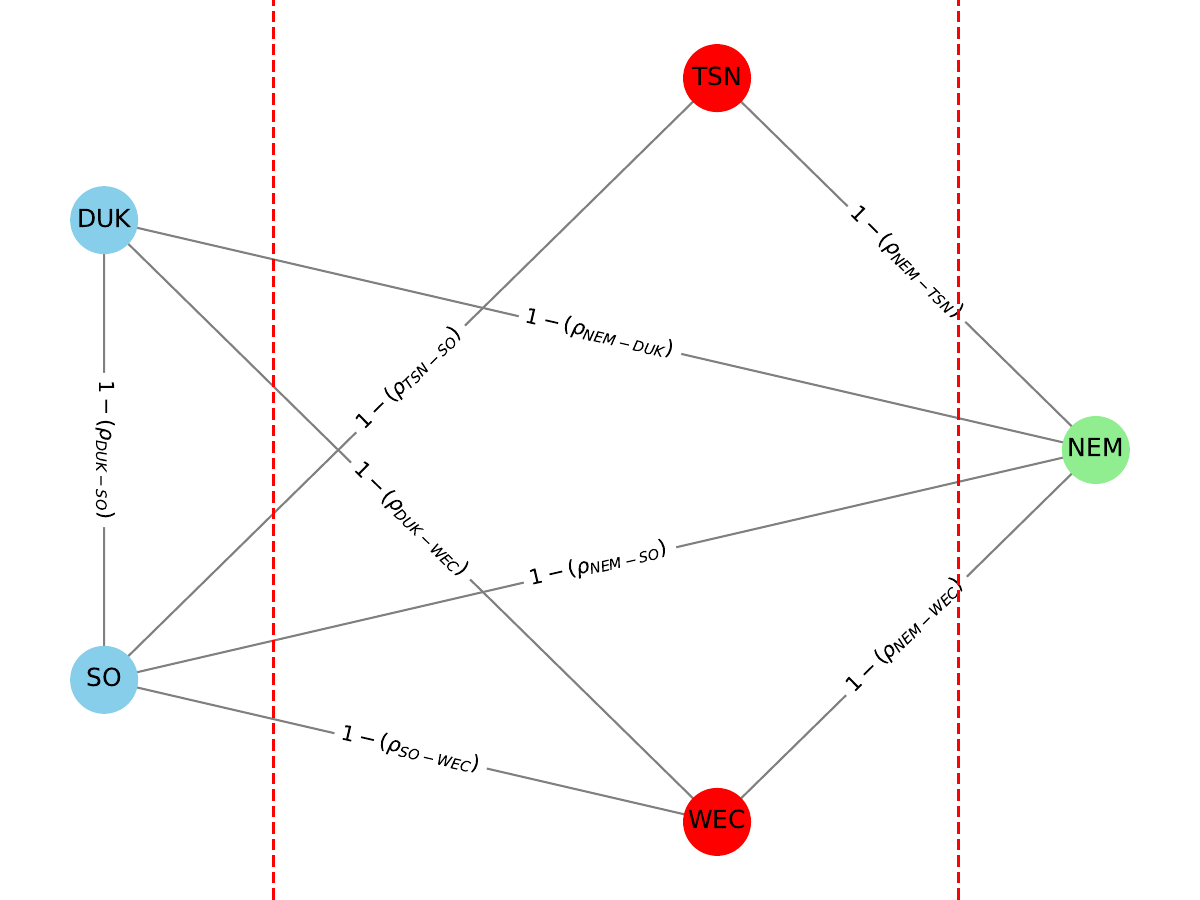}
\caption{Tripartition ($n_{splits} = 2$)}
\label{fig_graph_5_tripart}
\end{subfigure}
\caption{Iterative partition of the market graph (a) with $m=5$ nodes, into two (b) and three (c) subgraphs by performing cuts over the representation. Cuts are represented with red dashed lines, and same color nodes represent correlated assets within the same cluster.}
\label{fig_plain_bi_tri}
\end{figure*}

The optimal choice of cuts over $\mathcal{G}$, will find the optimal portfolio for the desired strategy of investment. This graph-based approach has been traditionally exploited for portfolio diversification from the classical computation perspective for multiple purposes such as dynamic optimization \citep{arroyo2022dynamic}, risk analysis \citep{ciciretti2024network} and community detection \citep{zhao2021community}; but also from the quantum computing perspective for portfolio diversification \citep{acharya2025decomposition, soloviev2025scaling}.

In this paper, our experiments show results for graphs sizes of $m \in \{10, 20, 30, 50, 100, 150, 200, 250\}$. Figure~\ref{fig_graph_10_20_30_plain} shows the graph representations for $m \in \{10, 20, 30\}$ and Table~\ref{tab_graph_info} summarizes the structural characteristics of market graphs for the interest of visualization. For each graph size, the number of edges, graph density, average node degree, and average clustering coefficient are shown. These metrics provide insight into the connectivity and local structure of the graphs. As the number of nodes increases, both the average degree and the clustering coefficient tend to grow, indicating a denser and more interconnected topology. The density remains relatively stable across scales. While some works \citep{parekh2025no, stopfer2025quantum} state the efficiency of classical solvers for sparse graphs, our study focuses on dense market graphs (see as shown in table), where classical cut-finding becomes computationally demanding and quantum methods may offer practical advantages.

\subsection{Bipartition strategy}
\label{sec_bipart_strategy}
Building upon the market graph $\mathcal{G}$ introduced in the previous Section, we propose a recursive bipartitioning methodology aimed at decomposing the graph into a predefined number of subgraphs. The core principle of this approach is to iteratively split $\mathcal{G}$ into two disjoint subsets by optimizing a cost function that reflects the structural dependencies among assets. Specifically, the algorithm seeks to maximize the number of cuts performed, thereby ensuring that each resulting subgraphs comprises assets that are highly correlated within each subgraph. 

To formalize this, we define the following cut function:
\begin{equation}
\label{eq_max_cut}
    \text{Cut}(\mathcal{G}, \bm{x}) = \sum_{v_i, v_j \in E} w_{ij}x_i(1 - x_j),
\end{equation}
where $\bm{x} \in \{0,1\}^m$ is a binary vector of length $m = |V(\mathcal{G})|$. Each entry $x_i$ denotes the assignment of asset $i$ to one of the two subgraphs: $x_i = 0$ if asset $i$ belongs to subgraph $\mathcal{G}_0$, and $x_i = 1$ otherwise. The edge weights $w_{ij}$ are defined following Equation~\ref{eq_weights}.

By maximizing Equation~\ref{eq_max_cut}, the algorithm encourages the separation of weakly connected components while preserving strong intra-subgraph connectivity. This results in partitions that are internally cohesive and externally sparse, which is a desirable outcome for our approach. 

\begin{algorithm*}[t]
\caption{Generate $n$ Portfolio Partitions via PCE Bipartitioning}
\label{alg_n_partition_combined}
\KwIn{Number of splits $n_{\text{splits}}$, Graph $\mathcal{G}$, \texttt{measurement}, order $k$}
\KwOut{Partition $(\mathcal{L}_1, \dots, \mathcal{L}_{n_{\text{splits}} + 1})$}

Initialize $G_{\text{to\_split}} \gets [\mathcal{G}]$\;
Initialize $L_{\text{labels}} \gets [0]$\;

\SetKwFunction{Bipartition}{bipartition}
\SetKwProg{Fn}{Function}{:}{}
\Fn{\Bipartition{$\mathcal{G}$, \texttt{measurement}, $k$}}{
    $\bm{\theta} \gets$ Initialize symbol map\;
    $n \gets$ Find $n$ such that $m \leq 3\binom{n}{k}$\;
    $p \gets \lfloor m / n \rfloor$\;
    $\Psi(\bm{\theta}) \gets$ Prepare HEA circuit with $n$ qubits and $p$ layers\;

    \SetKwFunction{Evaluate}{f}
    \SetKwProg{FnInner}{Function}{:}{}
    \FnInner{\Evaluate{$\bm{\theta}$}}{
        $\Psi \gets$ Set circuit parameters to $\bm{\theta}$\;
        $\bm{\Pi^k} = \{\Pi_1^k, \dots, \Pi_m^k\} \gets$ Compute $\bra{\Psi} \Pi_i \ket{\Psi}$ for all $i \in [1, m]$ using \texttt{measurement}\;
        $\bm{x} = [x_1, \dots, x_m] \gets$ Compute binary representation (Eq.~\ref{eq_sgn_func})\;
        Compute loss function $\mathcal{L}(\mathcal{G}, \bm{\Pi^k})$ (Eq.~\ref{eq_loss})\;
        \Return{$(\mathcal{L}(\mathcal{G}, \bm{\Pi^k}), \bm{x})$}
    }

    $(\mathcal{L}_{\min}, \bm{x}_{\min}) \gets$ \texttt{minimize} \Evaluate using classical optimizer\;
    $S_1 \gets \{i \mid \bm{x}_{\min}[i] > 0\}$\;
    $S_2 \gets \{i \mid \bm{x}_{\min}[i] \leq 0\}$\;
    \Return{$(S_1, S_2)$}
}
$ $

\For{$i \gets 1$ \KwTo $n_{\text{splits}}$}{
    $G_{\text{aux}} \gets G_{\text{to\_split}}[0]$\;
    \If{$|V(G_{\text{aux}})| > 1$}{
        $S_1, S_2 \gets$ \Bipartition{$G_{\text{aux}}$, \texttt{statevector}, $k$}\;
        Append $G[S_1]$ and $G[S_2]$ to $G_{\text{to\_split}}$\;
        Append $S_1$ and $S_2$ to $L_{\text{labels}}$\;
    }
    \Else{
        Append $G_{\text{to\_split}}[0]$ to $G_{\text{to\_split}}$\;
        Append $L_{\text{labels}}[0]$ to $L_{\text{labels}}$\;
        $n_{\text{splits}} \gets n_{\text{splits}} + 1$\;
    }
    Remove first element from $G_{\text{to\_split}}$\;
    Remove first element from $L_{\text{labels}}$\;
}
\Return{$L_{\text{labels}}$}
\end{algorithm*}

Algorithm~\ref{alg_n_partition_combined} describes the proposed strategy, where by performing $n_{splits}$ splits, $n_{splits} + 1$ subgraphs are generated. The algorithm iteratively calls $\texttt{bipartition}(G_{\text{aux}}, \texttt{measurement}, k)$ (line 22) over graph $G_{\text{aux}}$ from which two subsets of assets are returned. The algorithm iteratively splits the subgraphs until the convergence criteria is met (line 19). In our experiments we use the \texttt{statevector} simulation as the measurement in the approach, although it is generalizable to other types of shot-based measurements, as suggested in \cite{sciorilli2025towards} for PCE approach. The function \texttt{bipartition} (line 3) executes the PCE approach (Section~\ref{sec_pce}) to solve the loss function (Equation~\ref{eq_max_cut}) and maps the resultant binary string into two subgraphs. This function internally computes the required number of qubits (line 5) and corresponding number of layers (line 6) for the HEA circuit (line 7). A classical optimizer is run to optimally tune the parameters (line 14). In this paper, the COBYLA optimizer \citep{gomez2013advances} was used.

Figure~\ref{fig_plain_bi_tri} shows a toy example of the partition approach where a real market graph with $m=5$ nodes (a) is split $n_{splits} = 1$ (b), and $n_{splits} = 2$ (c) times, respectively. Same color nodes represent highly correlated nodes within the same cluster of assets.

The outcome of Algorithm~\ref{alg_n_partition_combined} is a collection of $n_{\text{splits}} + 1$ disjoint clusters, each comprising a subset of highly correlated assets. These clusters, denoted as $\{L_1, L_2, \dots, L_{n_{\text{splits}} + 1}\}$, serve as the foundation for constructing a diversified portfolio. To ensure that the portfolio captures the most promising investment opportunities within each cluster, we select a single representative asset from each group according to its historical performance.

\begin{figure*}[t]
    \centering
    \includegraphics[width=\linewidth]{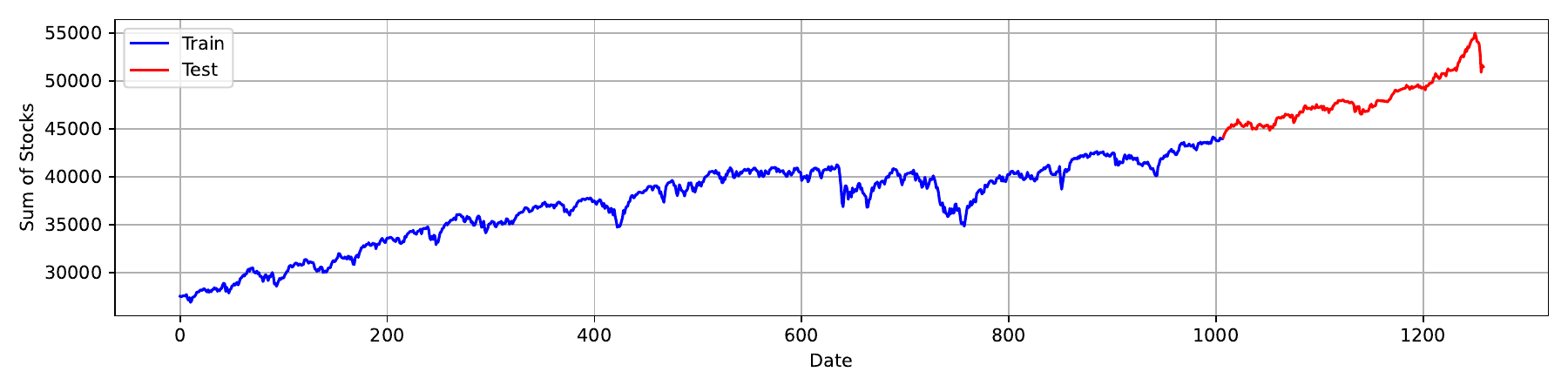}
    \caption{Aggregate of stock values in S\&P 500 stock market along time. Blue and red parts represent training and testing fractions of the dataframe, respectively.}
    \label{fig_sum_stocks}
\end{figure*}

Let $L_i = [L_i^1, \dots, L_i^{|L_i|}]$ denote the set of assets in cluster $L_i$, where $|L_i|$ is the cardinality of the cluster. For each asset $L_i^j$ in the cluster, we compute its return $\mu_i^j$ using a predefined return function (see Equation~\ref{eq_return}). This yields a return vector:
\[
\bm{\mu}_i = [\mu_i^1, \dots, \mu_i^{|L_i|}],
\]
which encapsulates the performance of all assets within cluster $L_i$. The representative asset $r_i$ for cluster $L_i$ is then selected as:
\[
r_i = \arg\max_{j \in \{1, \dots, |L_i|\}} \mu_i^j,
\]
i.e., the asset with the highest return within its respective cluster. This selection criterion ensures that the final portfolio is composed of top-performing assets from structurally coherent groups, thereby balancing diversification with return optimization. We aware here that other approaches have been used in the literature such as analyzing the sign of the pairwise correlation (\ref{eq_pearson_pair}) among the assets within the same cluster $L_i$ \citep{aref2019balance, harary2002signed}.

The resulting portfolio, composed of $\bm{r} = \{r_1, r_2, \dots, r_{n_{\text{splits}} + 1}\}$, is expected to exhibit reduced internal redundancy due to the high intra-cluster correlation and enhanced overall performance due to the return-based selection mechanism.

\subsection{Results}
\label{sec_results}
This section presents the experimental results obtained by applying the methodology outlined in Section~\ref{sec_bipart_strategy} to the problem instances described in Section~\ref{sec_market_graph_rep}.

The dataset was partitioned into training and testing subsets, comprising 80\% and 20\% of the original data, respectively. The partitioning strategy was applied to the training set to identify a subset $\bm{r}$ of representative assets. The performance of the selected assets was subsequently evaluated on the testing set. Figure~\ref{fig_sum_stocks} illustrates the temporal evolution of the aggregate value of all assets in the dataset, providing an overview of the dataset's dynamics over time, where blue and red parts represent the training and test fractions, respectively.

\begin{figure}[h]
    \centering
    \includegraphics[width=\linewidth]{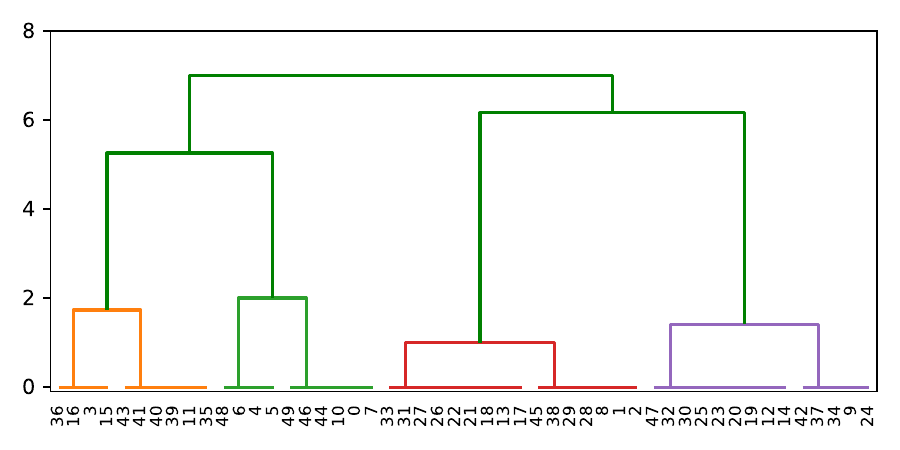}
    \caption{Dendogram for $m=50$, where each number in x-asis corresponds to a different stock company.}
    \label{fig_dendogram}
\end{figure}

As described previously, the iterative partitioning procedure recursively divides a subset of assets into two distinct subsets at each iteration. The dendrogram presented in Figure~\ref{fig_dendogram} visually captures the hierarchical structure resulting from the algorithm outlined in Algorithm~\ref{alg_n_partition_combined}. Each leaf node in the dendrogram corresponds to a group of highly correlated assets, denoted as $L_i$, for the case where $m = 50$. For clarity, individual assets have been anonymized and represented by numerical identifiers.

The vertical axis of the dendrogram reflects the depth of the recursive partitioning process, effectively mapping the sequence of iterations. Notably, the dendrogram reveals that the cardinality of the leaf-level groups $L_i$ remains relatively balanced across the entire set, a structural property that persists consistently across all tested values of $m$. This observation suggests that the partitioning algorithm tends to produce evenly sized clusters.

In this study, we assess the performance of our proposed method across different problem instances, with sizes ranging from $m = 10$ to $m = 250$, as summarized in Table~\ref{tab_graph_info}. To benchmark our approach against established optimization techniques, we consider two representative algorithms: QAOA, serving as a quantum method, and Estimation of Distribution Algorithms (EDAs), representing classical heuristics. QAOA is configured with $p = 5$ layers and applied exclusively to the smallest instance ($m = 10$), where the use of one-hot encoding within the partitioning framework remains computationally feasible. For larger instances ($m \geq 10$), EDAs are employed due to their scalability and adaptability to high-dimensional combinatorial spaces.

To ensure fair comparison across varying problem sizes, the number of recursive splits in the partitioning process is tuned according to instance size: specifically, $r_{\text{splits}} = {2, 4, 6, 9}$ for $m < 100$, and $r_{\text{splits}} = (m/10) - 1$ for $m \geq 100$.

\begin{figure*}[t]
\centering
\begin{subfigure}[b]{0.49\textwidth}
\centering
\includegraphics[width=\textwidth]{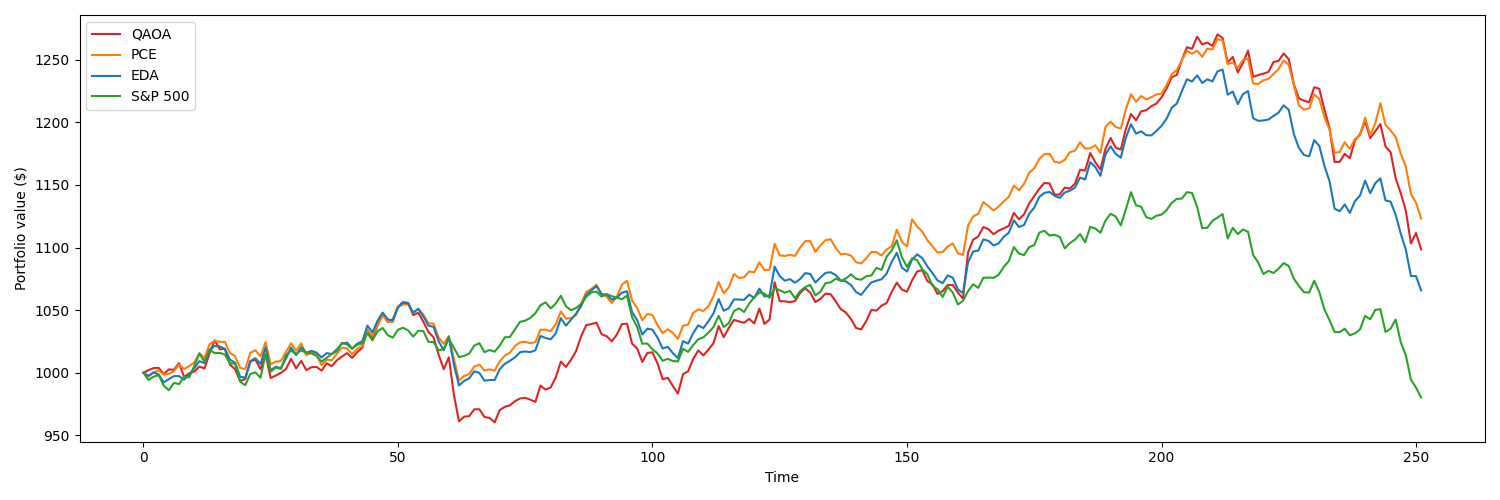}
\caption{$m=10$ and $n_{splits}=2$}
\label{fig:sub1}
\end{subfigure}
\begin{subfigure}[b]{0.49\textwidth}
\centering
\includegraphics[width=\textwidth]{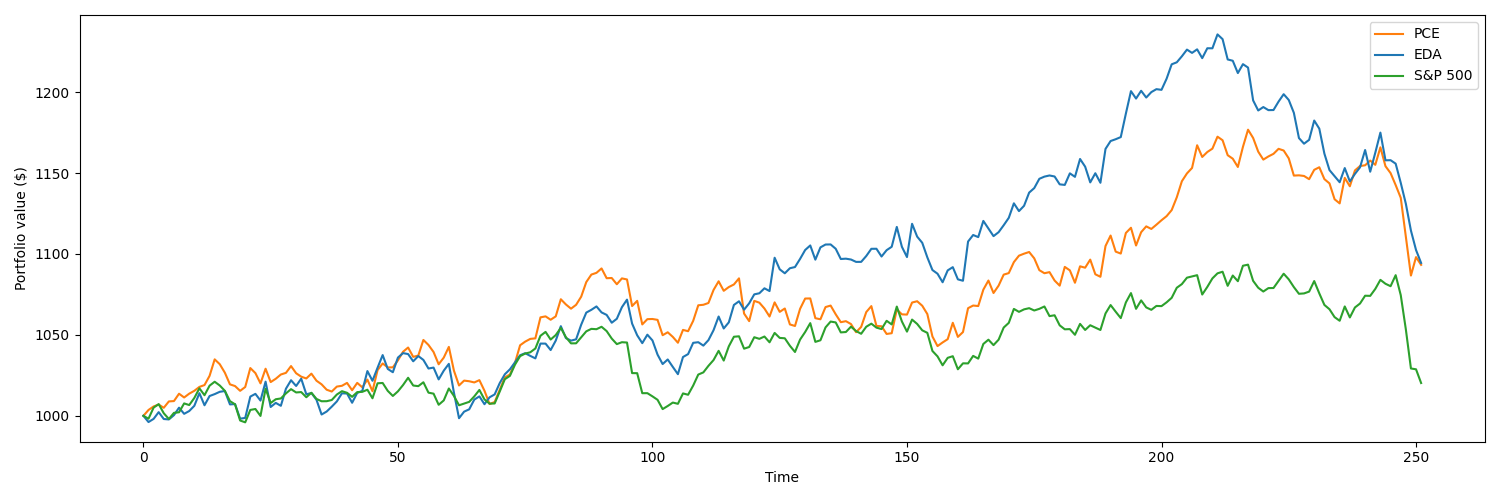}
\caption{$m=20$ and $n_{splits}=4$}
\label{fig:sub2}
\end{subfigure}
\begin{subfigure}[b]{0.49\textwidth}
\centering
\includegraphics[width=\textwidth]{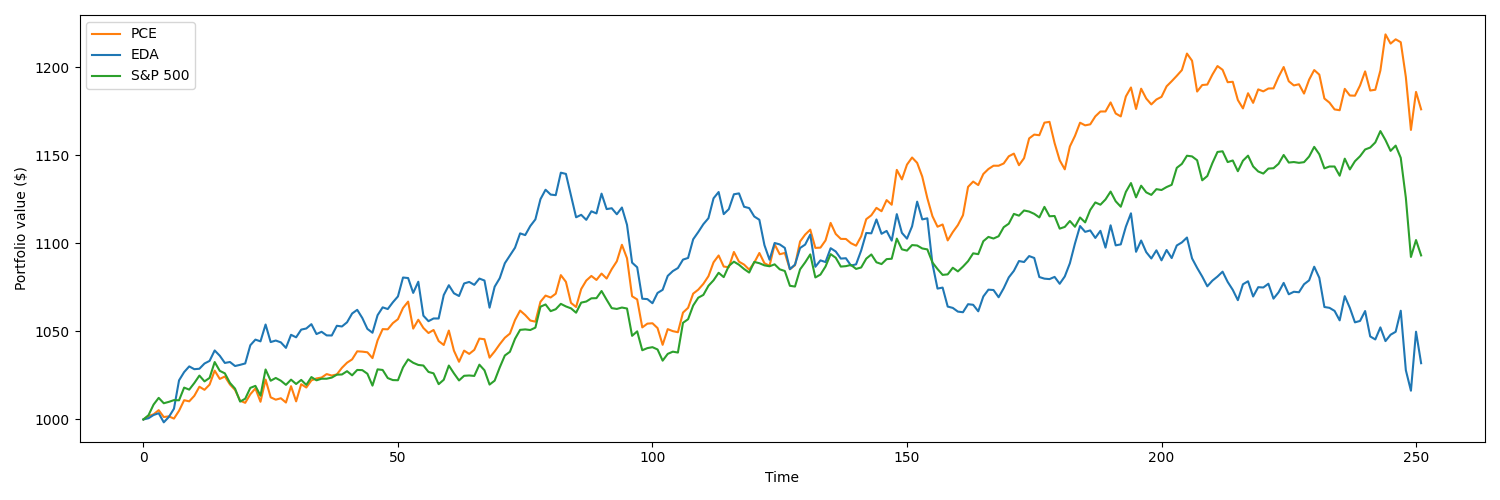}
\caption{$m=30$ and $n_{splits}=6$}
\label{fig:sub3}
\end{subfigure}
\begin{subfigure}[b]{0.49\textwidth}
\centering
\includegraphics[width=\textwidth]{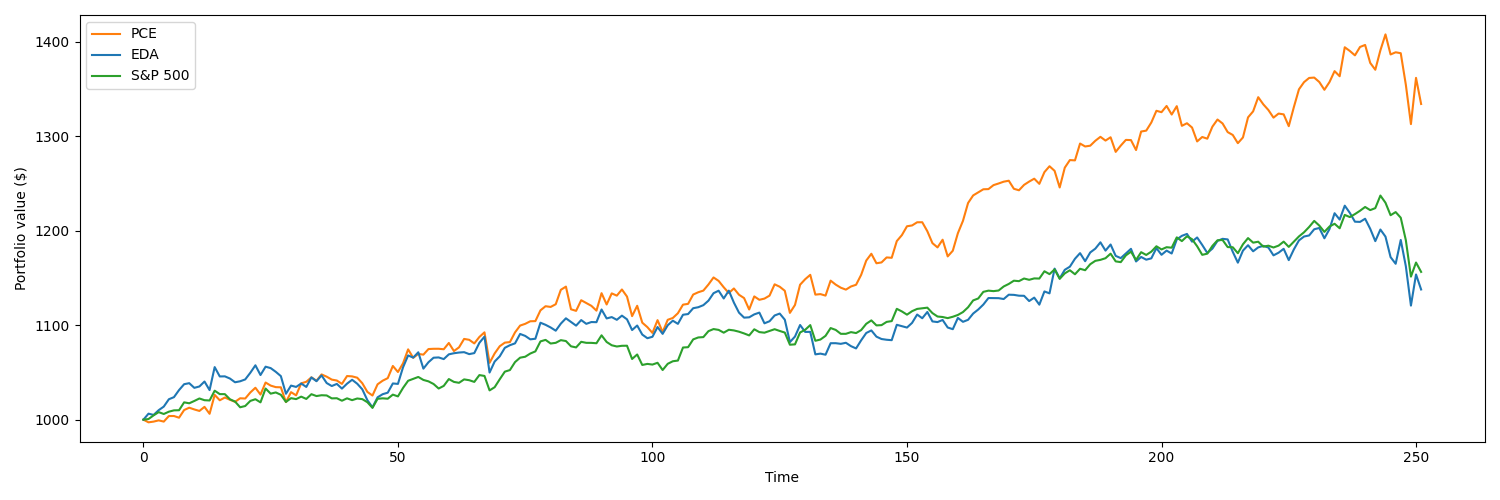}
\caption{$m=50$ and $n_{splits}=9$}
\label{fig:sub4}
\end{subfigure}
\begin{subfigure}[b]{0.49\textwidth}
\centering
\includegraphics[width=\textwidth]{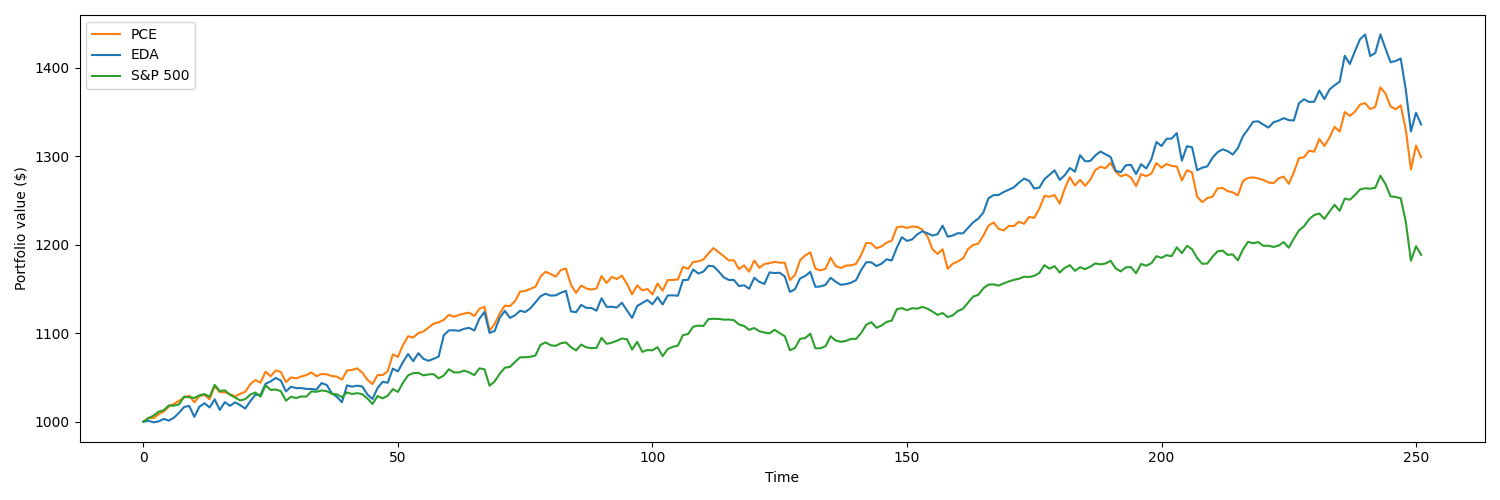}
\caption{$m=100$ and $n_{splits}=9$}
\label{fig:sub5}
\end{subfigure}
\begin{subfigure}[b]{0.49\textwidth}
\centering
\includegraphics[width=\textwidth]{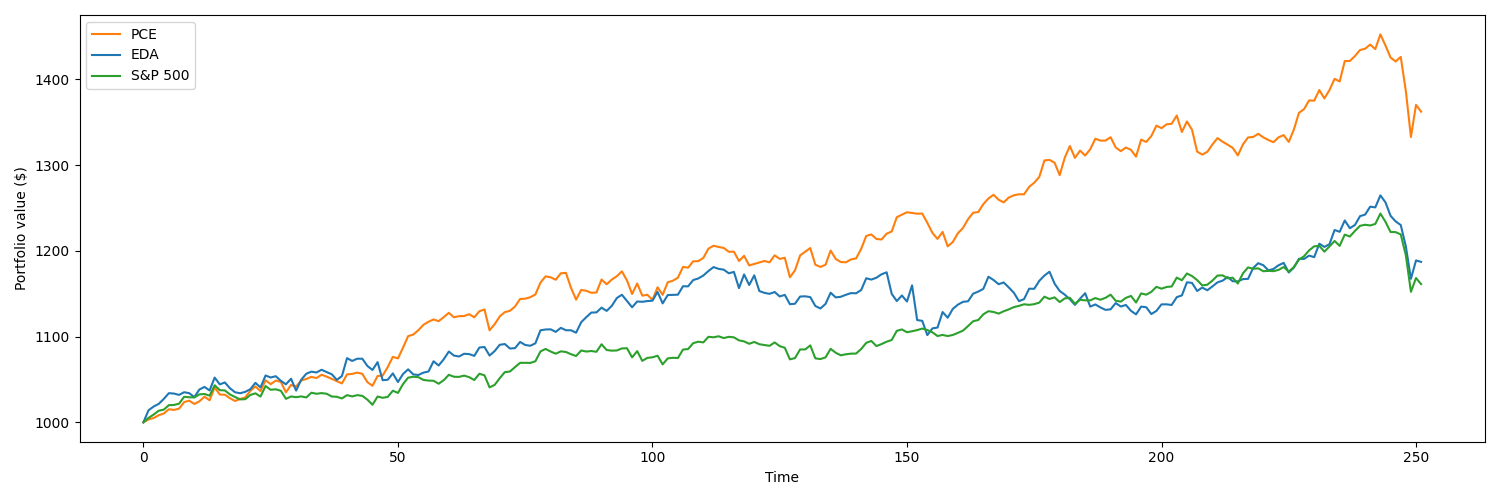}
\caption{$m=150$ and $n_{splits}=14$}
\label{fig:sub5}
\end{subfigure}
\begin{subfigure}[b]{0.49\textwidth}
\centering
\includegraphics[width=\textwidth]{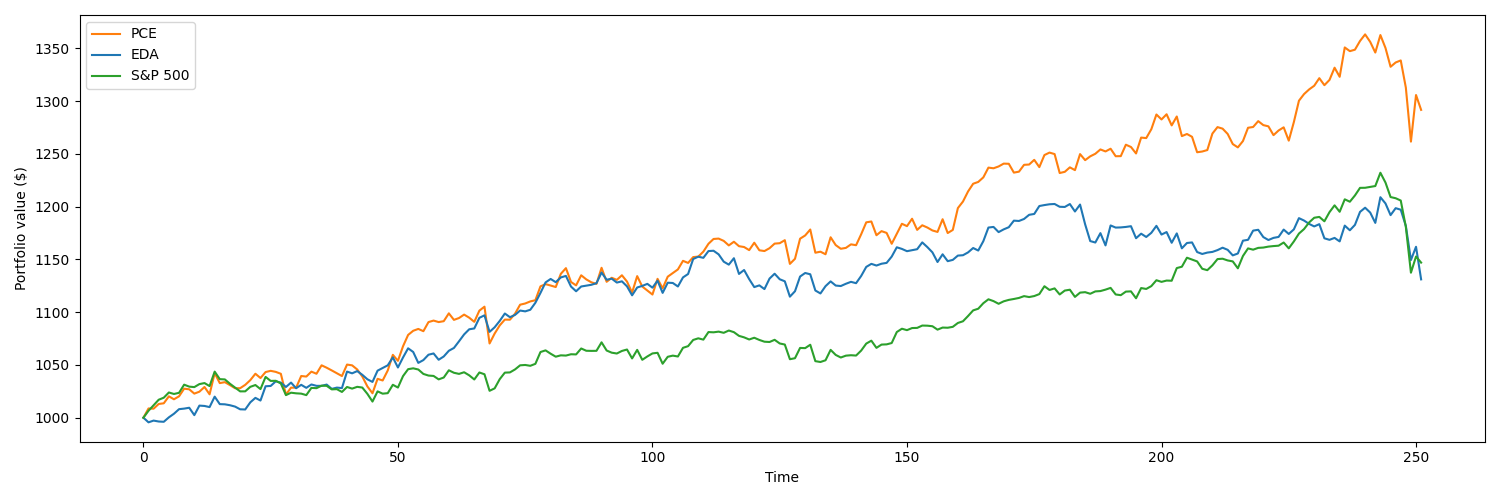}
\caption{$m=200$ and $n_{splits}=19$}
\label{fig:sub5}
\end{subfigure}
\begin{subfigure}[b]{0.49\textwidth}
\centering
\includegraphics[width=\textwidth]{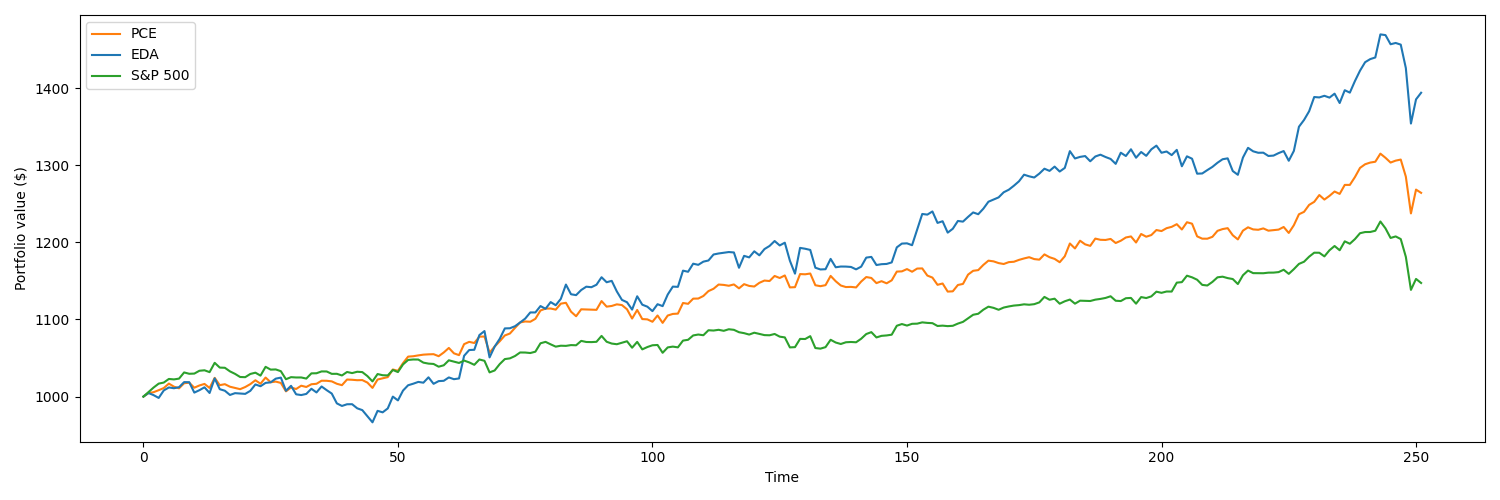}
\caption{$m=250$ and $n_{splits}=24$}
\label{fig:sub7}
\end{subfigure}
\caption{Comparison between money invested in S\&P 500 stock market against the portfolio optimization strategy proposed.}
\label{fig_benchmark_10_250}
\end{figure*}

Figure~\ref{fig_benchmark_10_250} presents the benchmarking results, where QAOA ($p = 5$), PCE ($k = 3$), and EDA are depicted as red, orange, and green lines, respectively. Each algorithm executes the procedure described in Algorithm~\ref{alg_n_partition_combined}, using historical data from the training set to identify an optimal subset of representative assets $\bm{r}$. The resulting portfolios are then evaluated on the test dataset, and their performance is visualized in the corresponding subplots. The figure illustrates the evolution of an initial investment of \$1000 over time, following the portfolio strategy derived from each optimizer, and compares it against the baseline performance of the full set of $m$ assets without optimization.

\begin{table*}[t]
    \centering
    \begin{tabularx}{\textwidth}{@{}X|X|X|X|X|X@{}}
        \toprule
        \textbf{\# Nodes} & \textbf{\# Edges} & QAOA ($p=1$) & QAOA ($p=2$) & PCE ($k=2$) & PCE ($k=3$) \\
        \midrule
        10  & 39    & 59    & 108   & 25   & 25   \\
        20  & 137   & 177   & 334   & 61   & 61   \\
        30  & 230   & 290   & 550   & 91   & 91   \\
        50  & 714   & 814   & 1578  & 148  & 145  \\
        100 & 3154  & 3354  & 6608  & 298  & 295  \\
        150 & 7606  & 7906  & 15662 & 430  & 433  \\
        200 & 14163 & 14563 & 28926 & 586  & 595  \\
        250 & 20948 & 21448 & 42646 & 715  & 730  \\
        \bottomrule
    \end{tabularx}
    \vspace{1em}
    \caption{Gate counts for QAOA (built with CZ) and PCE across graph sizes.}
    \label{tab_gate_counts}
\end{table*}

The performance for $m=10$ for QAOA and PCE is very similar in terms of the selection of assets. However, since QAOA is limited by the qubit encoding used, the problem instances tested might not be enough to conclude the performance of the approach. Table~\ref{tab_gate_counts} reports the gate counts required by QAOA and PCE algorithms across a range of graph sizes, characterized by the number of nodes and edges. The QAOA gate counts are shown for two circuit depths, $p = 1$ and $p = 2$, while PCE shows order $k = 2$ and $k = 3$. As expected, the gate complexity for QAOA increases significantly with both the number of nodes and the circuit depth, reflecting the quadratic scaling of entangling operations in dense graphs. For instance, QAOA with $p = 2$ requires over 42,000 gates for a graph with 250 nodes and nearly 21,000 edges. In contrast, PCE exhibits a much more modest growth in gate count, remaining under 750 gates even for the largest instance tested. This highlights the scalability advantage of PCE over QAOA, particularly for large problem instances where quantum resource constraints are prohibitive. Since $p=5$ has been chosen for this problem instance, the scalability of QAOA is limited by the depth and number of qubits used. 

For greater number of nodes $m \geq 10$, it is observed that the proposed strategy always outperforms the baseline for all the tested cases, regardless of the optimization procedure used. PCE shows the most robust and stable growth trajectory across all problem sizes. Its portfolio value steadily increases over time, often maintaining a clear lead over both EDA and the S\&P 500. This suggests that PCE is effective at identifying representative assets that contribute to long-term portfolio growth. EDA, while also outperforming the baseline, exhibits greater volatility. It occasionally reaches higher peaks than PCE ($m=100$ and $m=250$) but also experiences sharper drawdowns. This behavior indicates that while EDA can capture short-term opportunities, it may be more sensitive to market fluctuations or local optima in the asset selection process. The iterative behavior exhibited by the EDA reveals a distinct partitioning strategy, wherein a small number of nodes are incrementally separated from the main graph structure during each iteration. This contrasts with PCE, which consistently performs more balanced splits, typically dividing the graph into two subsets of comparable size. Such divergence in partitioning dynamics reflects fundamental differences in the underlying heuristics of each method and may influence the granularity and stability of $\bm{r}$ found.

\begin{figure}[h]
\centering
\begin{subfigure}[b]{0.49\textwidth}
\centering
\includegraphics[width=\textwidth]{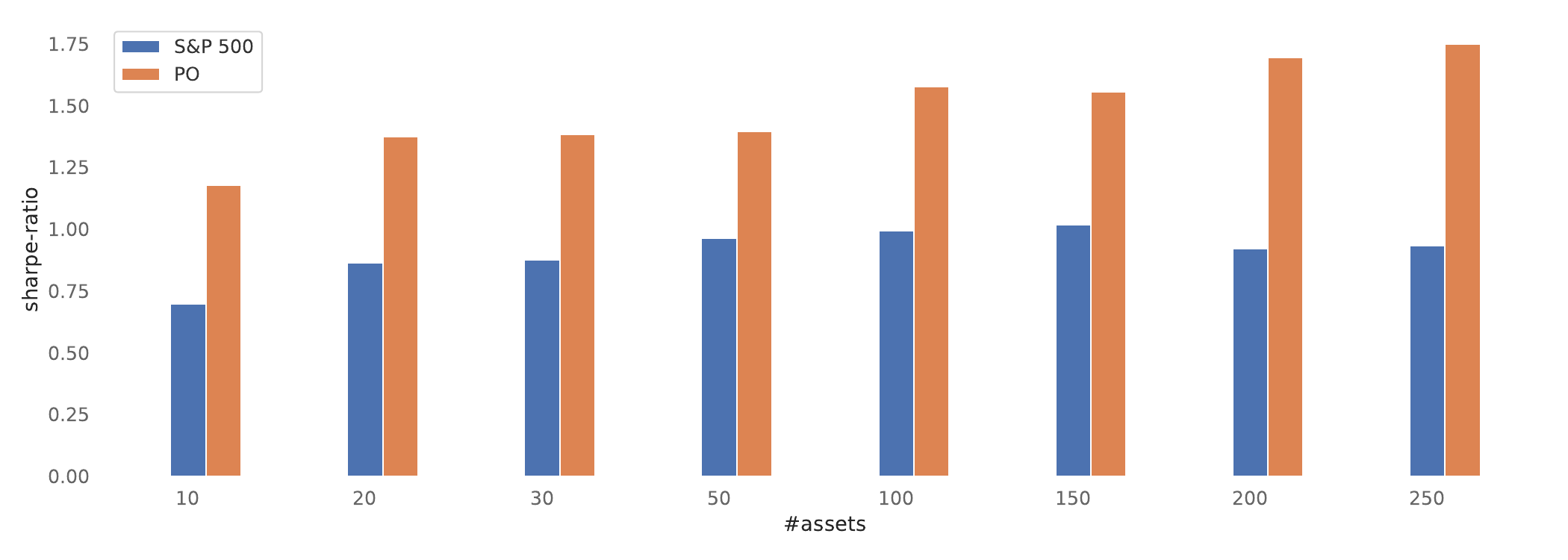}
\caption{Train}
\label{fig_sharpe_ratio_train}
\end{subfigure}
\begin{subfigure}[b]{0.49\textwidth}
\centering
\includegraphics[width=\textwidth]{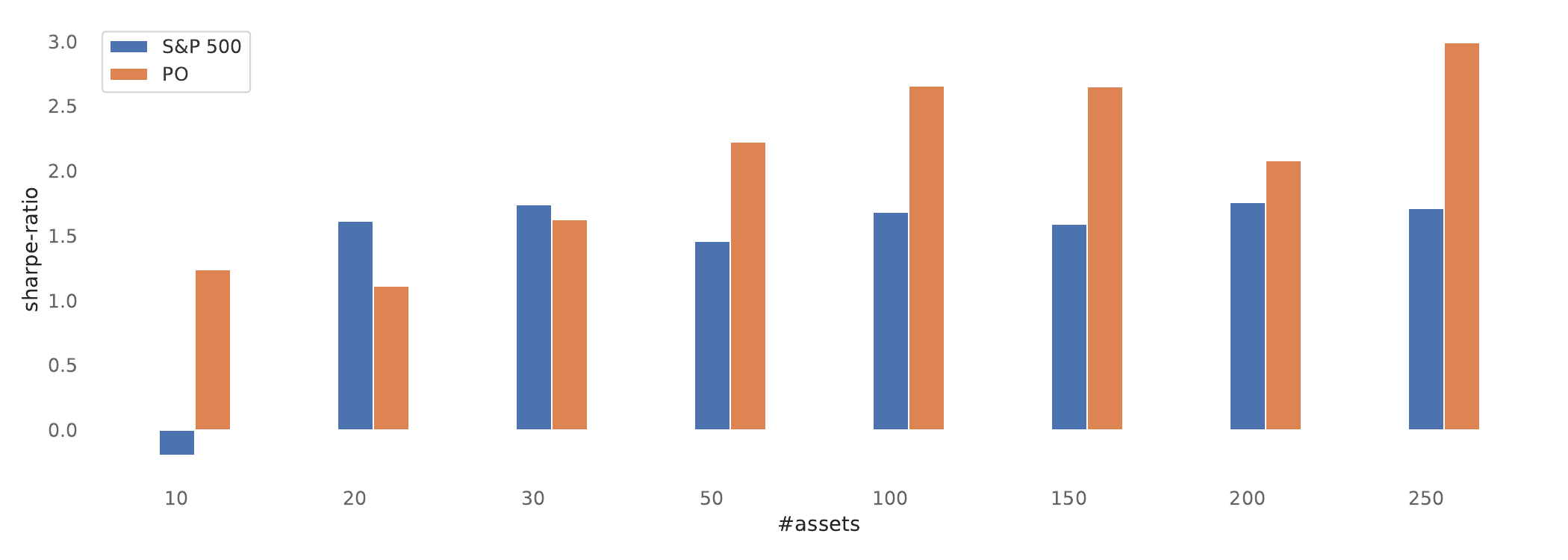}
\caption{Test}
\label{fig_sharpe_ratio_test}
\end{subfigure}
\caption{Sharpe-ratio comparison}
\label{fig_sharpe_ratio}
\end{figure}

Since the original market graph $\mathcal{G}$ encodes pairwise Pearson correlations among stock assets, the optimization strategy initially focuses on structural properties rather than asset returns. Consequently, the selection of representative assets $\bm{r}$ is performed over the decision made in the graph. To evaluate the effectiveness of the selected representatives, Figure~\ref{fig_sharpe_ratio} presents a comparative analysis of the Sharpe ratio (Equation~\ref{eq_sharpe}) for the PCE strategy and the baseline, shown as orange and blue bars, respectively, across both training and test datasets.

Figure~\ref{fig_sharpe_ratio_train} shows that the PCE approach successfully identifies a subset $\bm{r}$ that yields a higher Sharpe ratio than the baseline, indicating superior risk-adjusted performance. This trend is consistent in the test dataset, as shown in Figure~\ref{fig_sharpe_ratio_test}, where PCE continues to outperform the baseline in most experiments. These results suggest that the proposed strategy effectively balances return maximization and risk minimization, as captured by the Sharpe ratio, despite its initial reliance on correlation structure rather than direct return information.

\begin{figure}[h]
    \centering
    \includegraphics[width=\linewidth]{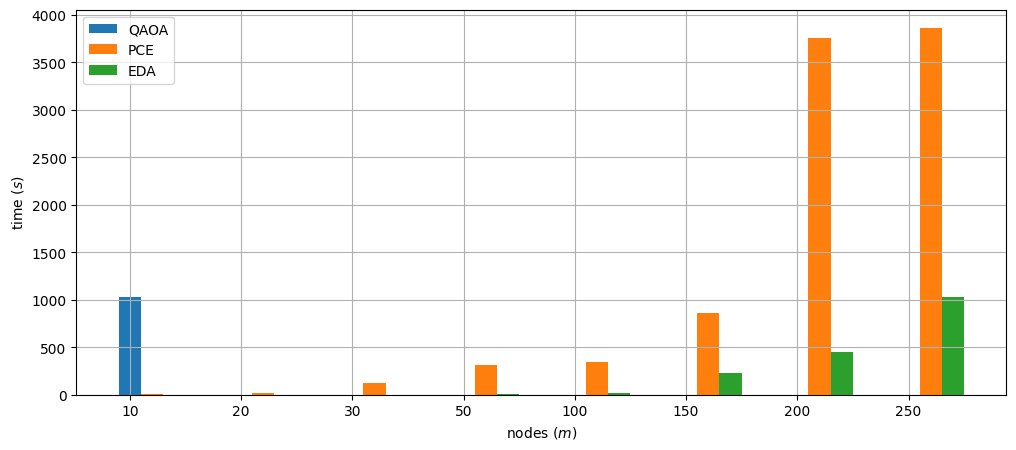}
    \caption{Times comparison (algorithm execution for single optimization problem)}
    \label{fig_times}
\end{figure}

To assess the computational efficiency of the proposed methods, Figure~\ref{fig_times} shows a runtime analysis for a single iteration of Algorithm~\ref{alg_n_partition_combined} with $n_{\text{splits}} = 1$. Figure~\ref{fig_times} presents the runtime as a function of the number of nodes ($m$) involved in the optimization. As illustrated, QAOA exhibits the highest computational cost, with its runtime for $m = 10$ comparable to that of the PCE approach for $m = 150$. Notably, PCE is capable of solving instances with $m \geq 200$ in approximately one hour, indicating its practical viability for real-world applications. Note that statevector simulation was used for these experiments, and this runtime may decrease if a different measurement strategy or a real quantum device were used. Although the classical heuristic (EDA) generally requires less computational effort, its runtime may vary depending on the configuration of its hyperparameters.

\section{Conclusions}
\label{sec_conclusions}

In this work, we proposed a graph-based portfolio optimization strategy that iteratively bipartitions a market graph representing the S\&P 500, where each node corresponds to a stock asset and edges encode pairwise Pearson correlations. From each identified group of correlated assets, a representative is selected to construct an optimized portfolio. Traditional quantum gate-based models such as QAOA face scalability limitations due to one-hot encoding, which maps each asset to a separate qubit. To address this, we propose the usage of PCE method, enabling the representation of multiple optimization variables per qubit.

Our results demonstrate that instances involving more than $m > 250$ assets can be solved within approximately one hour using a statevector simulator, surpassing the scalability constraints of conventional quantum approaches. The proposed strategy effectively balances return maximization and risk minimization, as evidenced by superior Sharpe ratio performance compared to baseline methods. Although this study scales up to $m = 250$ and $n=9$, the approach is designed to potentially accommodate the full S\&P 500 stock market or larger asset sets at a cost of increasing the number of qubits $n$.

While the current implementation of PCE utilizes a Hardware-Efficient Ansatz, future work include the exploration of alternative quantum circuit architectures tailored to this problem. Additionally, further research will investigate enhanced methods for selecting representative assets from each correlated group to improve portfolio construction. Finally, we also propose the analysis of quantum noise resilience as a further analysis.

\section*{Acknowledgments}
The authors would like to express their sincere gratitude to Bibhas Adhikari from Fujitsu Research of America for reviewing the paper and providing valuable feedback.

\bibliography{references}

\end{document}